# Preparation and ferroelectric properties of (124)-oriented $SrBi_4Ti_4O_{15}$ ferroelectric thin film on (110)-oriented $LaNiO_3$ electrode

*Xi Wang, Pilong Wang, Guangda Hu[a], Jing Yan, Xuemei Chen, Yanxia Ding, and Weibing Wu*

School of Materials Science and Engineering, University of Jinan, Jinan 250022, China

*Suhua Fan*

College of Materials Science and Engineering, Shandong Jianzhu University, Jinan 250101, China

___________________________________

a) Corresponding author, E-mail: mse_hugd@ujn.edu.cn

**Abstract**

A (124)-oriented $SrBi_4Ti_4O_{15}$ (SBTi) ferroelectric thin film with high volume fraction of $\alpha^{SBTi}_{(124)}$=97% was obtained using a metal organic decomposition process on $SiO_2$/Si substrate coated by (110)-oriented $LaNiO_3$ (LNO) thin film. The remanent polarization ($P_r$) and coercive field ($E_c$) for (124)-oriented SBTi film are 12.1 μC/cm$^2$ and 74 kV/cm, respectively. No evident fatigue of (124)-oriented SBTi thin film can be observed after $1\times10^9$ switching cycles. Besides, the (124)-oriented SBTi film can be uniformly polarized over large areas using a piezoelectric-mode atomic force microscope. Considering that the annealing temperature was 650°C and the thickness of each deposited layer was merely 30 nm, a long-range epitaxial relationship between SBTi(124) and LNO(110) facets was proposed. The epitaxial relationship was demonstrated based on the crystal structures of SBTi and LNO.

# 1. Introduction

In recent years, bismuth layer-structured ferroelectrics (BLSFs) have attracted much attention because they are considered to be the excellent candidates for applications in ferroelectric random access memories (FeRAMs) and micro-electromechanical systems (MEMS) [1-3]. The general formula of BLSFs is $(Bi_2O_2)^{2+}(A_{n-1}B_nO_{3n+1})^{2-}$, where *A* and *B* are metal ions, *n* is the the Aurivillius parameter corresponding to the number of perovskite layer. Among them, $SrBi_2Ta_2O_9$ (SBTa) (*n*=2) and lanthanide doped $Bi_4Ti_3O_{12}$ (BTO) (*n*=3) have been extensively studied so far [4,5]. Recently, much attention has been paid on 4-layer perovskite BLSFs (*n*=4) such as $SrBi_4Ti_4O_{15}$ (SBTi), $CaBi_4Ti_4O_{15}$ (CBTi) and $BaBi_4Ti_4O_{15}$ (BBTi) due to their good ferroelectric properties [6-8].

As a conductive metal oxide, $LaNiO_3$ (LNO) is a promising electrode material for ferroelectric capacitors due to its low resistivity, good thermal stability and better lattice match with most of ferroelectric thin films [9,10]. (100)-oriented LNO oxide electrodes have been used successfully to control the structure of SBTa [11], BTO [12] and BBTi [8] thin films. Satyalakshmi *et al.* found that c-axis-predominant BBTi thin films can be formed on (100)-oriented LNO [8]. This should result from the epitaxial relationship between BBTi(001) (a=0.386 nm, c=4.18 nm) and LNO(100) (a=b=0.384 nm). It is well known that the major polarization vector of 4-layer perovskite BLSFs is along the a axis, the growth of non-c-axis-oriented thin films is required to achieve a higher polarization component along the surface normal of the ferroelectric thin films. Considering the direct epitaxial relationship between (100)-oriented LNO and 4-layer perovskite BLSFs, (110)-oriented LNO film should be a better choice for the fabrication of non-c-axis-oriented 4-layer perovskite BLSFs thin films in comparison with (100)-oriented LNO film.

Up to now, few ferroelectric thin films have been prepared on (110)-oriented LNO electrodes [13].

In our previous work, we have deposited (100) and (110)-oriented LNO films on $SiO_2$/Si substrates using a metal organic decomposition (MOD) method [14]. In this work, we report on the preparation and ferroelectric properties of (124)-oriented SBTi ferroelectric thin films on (110)-oriented LNO electrode at a relative low annealing temperature of 650°C. Based on the crystal structures of SBTi and LNO, a long-range epitaxial relationship between the (124) facet of the former and the (110) plane of the latter is proposed.

## 2. Experimental

LNO(110) thin film was deposited on $SiO_2$/Si substrate using a metal organic decomposition (MOD) process, which has been described elsewhere [14]. MOD process was also used to fabricate SBTi films on LNO(110)/$SiO_2$/Si substrate. The precursor solution was prepared by dissolving bismuth nitrate, strontium nitrate, and titanium butoxide in an ethylene glycol. Acetylacetone was added to stabilize the solution. Bismuth of 10 mol % excess was added to compensate the bismuth loss during the annealing process. The film was deposited onto the substrate by spin coating. Each layer (about 30 nm) of the wet film was pre-annealed at 400°C for 10 min and subsequently annealed at 650°C for 8 min in a rapid thermal annealing furnace in an oxygen atmosphere. This step was repeated several times to obtain the desired thickness. Au top electrodes were deposited on SBTi film using a sputtering system through a shadow mask with a diameter of 0.2 mm for the electrical measurements.

The structure of LNO and SBTi films was studied by x-ray diffraction (XRD) using a Bruker D8 diffractometer with Cu *Kα* radiation. The surface morphologies and microscopic domain switching property of the SBTi film were detected by an atomic force microscope (AFM) in the piezoresponse mode. A standard ferroelectric tester (Precision Pro. Radiant Technologies) was used to measure the ferroelectric properties of the film.

## 3. Results and discussion

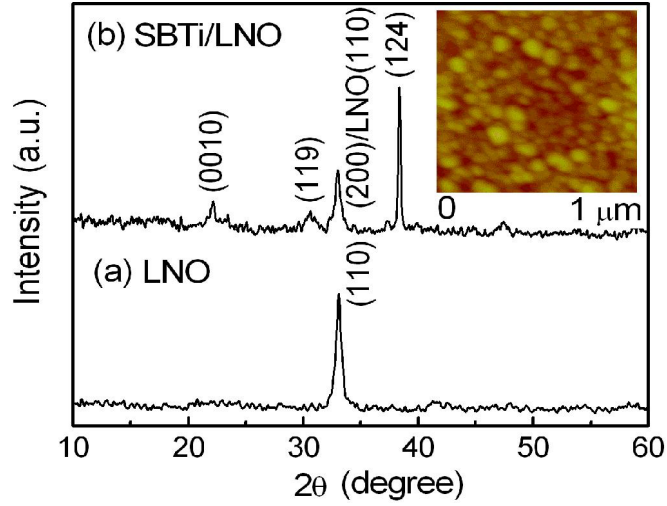

Fig.1 XRD pattern for (a) (110)-oriented LNO film and (b) the SBTi film deposited on LNO(110) electrode. The inset shows the AFM image of (110)-oriented LNO film.

Fig.1(a) shows the XRD pattern of LNO film prepared at 700°C. The intensity of (110) peak is much higher than those of (100) and (200) peaks. The value of volume fraction $\alpha$ for the (110)-oriented LNO grains is calculated to be more than 90%, based on the database of JCPDS. As shown in the inset of Fig.1, the surface of the (110)-oriented LNO is very smooth with a roughness value less than 1 nm.

Fig.1(b) illustrates the XRD pattern of the SBTi film deposited on LNO(110) electrode. The diffraction peak of (124) is obviously preferred. The full width at half maximum (FWHM) of the (124) peak is about 0.24°, indicating that the film is well crystallized. To evaluate the degree of the preferred orientation of (124), the volume fraction $\alpha$ is calculated using the equation:[15]

$$\alpha_{124} = (I_{124}/I^*_{124})/\Sigma \ (I_{hkl}/I^*_{hkl}) \qquad (1)$$

where $I_{hkl}$ is the measured intensity of the *(hkl)* peak for the film and $I^*_{hkl}$ is the intensity for the powder. The $\alpha$ value for the (124) peak of SBTi thin film is as high as 97%. To our knowledge, there is no report on the fabrication of (124)-oriented SBTi thin film so far. Considering that the layer thickness of the (124)-oriented SBTi film is about 30 nm during the layer-by-layer annealing process, the interface energy between SBTi film and bottom LNO electrode will have a strong effect on the structure of SBTi film. The effect of surface energy on the structure of SBTi film is also considered. In general, at a relative low temperature (e.g. ⩽650°C), the influence of the minimization of the interface energy on the crystal orientation of the film is greater than that of the surface energy minimization. Since the annealing temperature adopted in the present work is only 650°C, the effect of the interface energy minimization should be dominated for (124)-oriented SBTi film deposited on (110)-oriented LNO film. Therefore, there should exist an epitaxial relationship between SBTi(124) and LNO(110) facets.

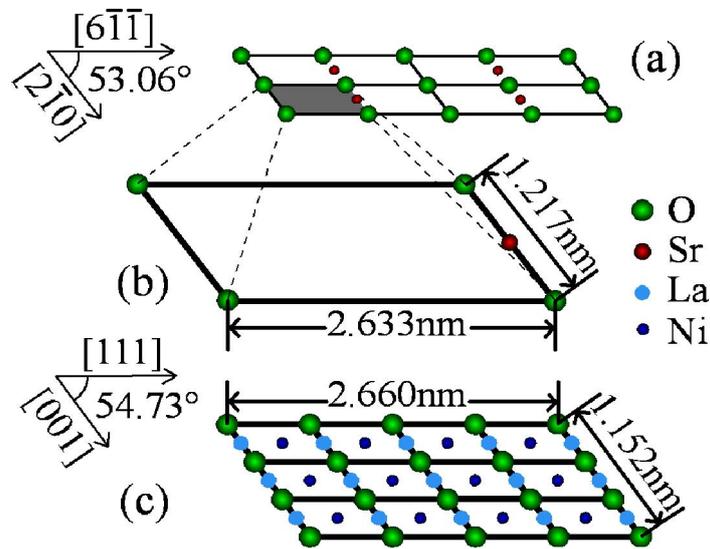

Fig. 2. Diagrams of the possible lattice matching relationship between the (124) facet of SBTi and the (110) plane of LNO.

To prove this hypothesis, the crystal structures of SBTi (orthorhombic structure with the lattice parameters of a=0.5445 nm, b=0.5437 nm, and c=4.095 nm) and LNO (cubic structure with the lattice parameter of a=0.384 nm) are analyzed and calculated. Fig.2(a) and (c) show the structure of (124) facet of SBTi and (110) plane of LNO, respectively. Fig.2(b) shows an amplified one of the grey area in Fig.2(a), in which the oxygen irons are the closest ones along $[6\bar{1}\bar{1}]$ and $[2\bar{1}0]$ in (124) plane of SBTi. For the case of SBTi, the angle between $[6\bar{1}\bar{1}]$ and $[2\bar{1}0]$, i.e., $\angle[6\bar{1}\bar{1}],[2\bar{1}0]$ is 53.06°, which is close to the angle of $\angle[111],[001]$ (54.73°) for LNO in Fig.2(c). The corresponding orientations of SBTi and LNO can be considered to be parallel to each other (i.e., SBTi$[6\bar{1}\bar{1}]\parallel$LNO$[111]$; SBTi$[2\bar{1}0]\parallel$LNO$[001]$), since small angular deviation less than 2° is tolerated due to the fact that the first deposited 30-nm-thick layer can be easily distorted. The interval of the closest oxygen ions along SBTi$[6\bar{1}\bar{1}]$ and the fourfold interval of the closest oxygen ions along LNO$[111]$ are about 2.633 nm and 2.660 nm, respectively. On the other hand, the interval of the closest oxygen ions along SBTi$[2\bar{1}0]$ and the threefold interval of the closest oxygen ions along LNO$[001]$ are about 1.217 nm and 1.152 nm, respectively. The lattice mismatches along SBTi$[6\bar{1}\bar{1}]\parallel$LNO$[111]$ and SBTi$[2\bar{1}0]\parallel$LNO$[001]$ are about 1% and 5.34%, respectively. These can be totally tolerated also due to the fact that the first deposited layer is very thin. Therefore, the epitaxial relationship between SBTi(124) and LNO(110) can be described as SBTi(124)$\parallel$LNO(110); SBTi$[6\bar{1}\bar{1}]\parallel$LNO$[111]$; SBTi$[2\bar{1}0]\parallel$LNO$[001]$. In comparison with the direct epitaxial relationship between (100)-oriented LNO and c-axis-oriented BBTi reported by Satyalakshmi *et al*.[8], the long-range epitaxial growth of (124)-oriented SBTi in the present work is more difficult to realize. We believe that the layer-by-layer process plays a key role in forming (124)-oriented SBTi film on (110)-oriented LNO oxide electrode owing to its flexibility on controlling the layer thickness and temperatures of heating treatments in each layer.

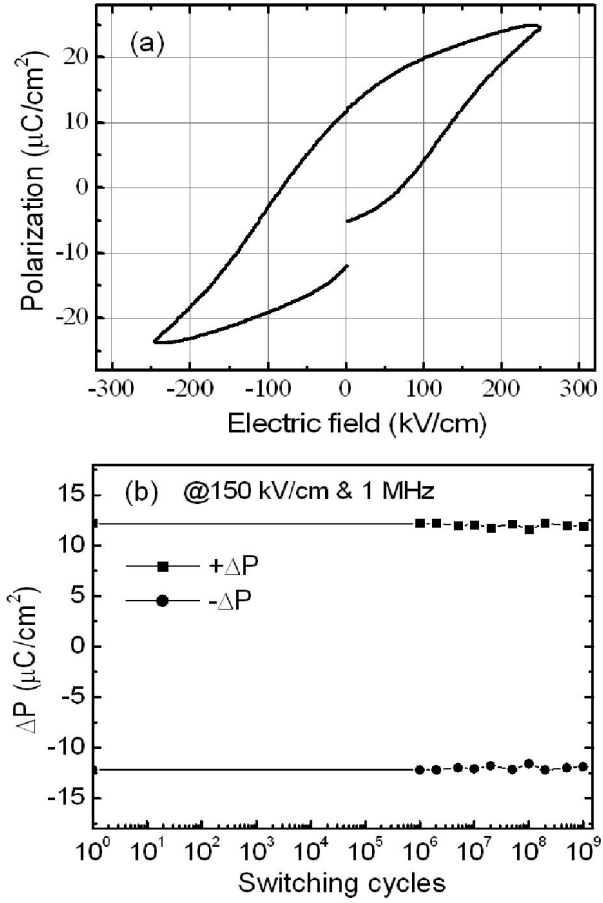

Fig. 3 (a) P-E hysteresis loops of (124)-oriented SBTi thin film and

(b) polarization fatigue of the Au/ SBTi(124)/LNO(110) capacitor.

Fig.3(a) shows the ferroelectric hysteresis loop of (124)-oriented SBTi film under an electric field of 250 kV/cm. The remanent polarization ($P_r$) and coercive field ($E_c$) for (124)-oriented SBTi film are 12.1 μC/cm$^2$ and 74 kV/cm, respectively. As shown in Fig.3(b), no evident fatigue can be observed after $1\times10^9$ switching cycles at 150 kV/cm amplitude and 1MHz frequency. Evidently, the fatigue-free property of SBTi film should be due to the fact that the LNO oxide electrode can effectively eliminate the oxygen vacancies located at the interface between SBTi and LNO films.

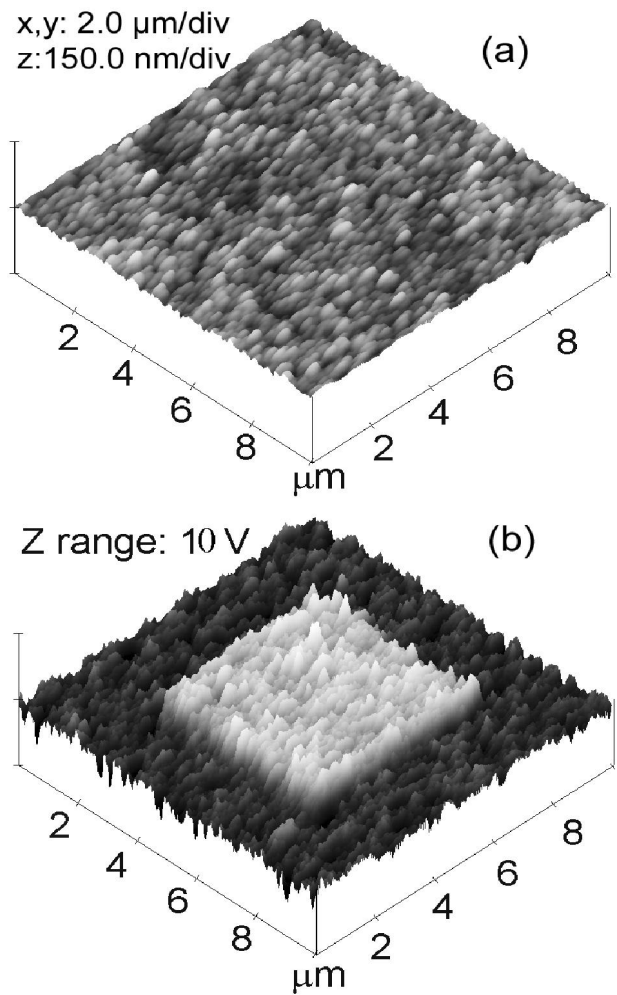

Fig.4 Simultaneously obtained (a) topographic and (b) domain images of the polarized (124)-oriented SBTi thin film deposited on LNO(110)/SiO$_2$/Si substrate.

Fig.4 shows the simultaneously obtained topographic and domain images of the polarized (124)-oriented SBTi thin film. The average surface roughness for (124)-oriented SBTi film is less than 6 nm, and the grain size is in the range of 100-150 nm. In the domain image shown in Fig.4(b), the bright square region (5μm×5μm) and the dark region (10μm×10μm) were formed by applying +10 and -10V voltage between the tip and bottom electrode, respectively. It can be seen that the SBTi thin film can be uniformly polarized and imaged without being affected by the film roughness. This provides further evidence that the SBTi film deposited on LNO electrode is highly (124)-oriented.

## 4. Conclusion

A (124)-oriented SBTi thin film with high volume fraction of $\alpha^{SBTi}_{(124)}$=97% was epitaxially grown on SiO$_2$/Si substrate coated by (110)-oriented LNO thin film. The long-range epitaxial relationship between SBTi(124) and LNO(110) can be described as SBTi(124) $\|$ LNO(110); SBTi[$6\bar{1}\bar{1}$] $\|$ LNO[111]; SBTi[$2\bar{1}0$] $\|$ LNO[001]. The remanent polarization ($P_r$) and coercive field ($E_c$) for (124)-oriented SBTi film are 12.1 μC/cm$^2$ and 74 kV/cm, respectively. No evident fatigue can be observed after $1\times10^9$ switching cycles. Besides, the (124)-oriented SBTi film is quite smooth and can be uniformly polarized over large areas using a piezoelectric-mode atomic force microscope.

## Acknowledgements

This work was supported by funding from National Natural Science Foundation of China (Grant No. 90207025 and 50502016) and the Natural Science Foundation of Shandong Province, China.

## References

[1] J. F. SCOTT and C. A. PAZ DE ARAUJO, Science **246** (1989) 1400.

[2] B. H. PARK, B. S. KANG, S. D. BU, T. W. NOH, J. LEE and W. JO, Nature **401** (1999) 682.

[3] C. A. PAZ DE ARAUJO, J. D. CUCHIARO, L. D. MCMILLAN, M. C. SCOTT and J. F. SCOTT, Nature(London) **374** (1995) 627.

[4] H. TABATA, H. TANAKA and T. KAWAI, Jpn. J. Appl. Phys. **34** (1995) 5146.

[5] T. LI, Y. ZHU, S. B. DESU, C. H. PENG and M. NAGATA, Appl. Phys. Lett. **68** (1996) 616.

[6] D. S. SOHN, W. X. XIANYU, W. I. LEE, I. LEE and I. CHUNG, Appl. Phys. Lett. **79** (2001) 3672.

[7] K. KATO, K. SUZUKI, K. NISHIZAWA and T. MIKI, J. Appl. Phys. **89** (2001) 5088.

[8] K. M. SATYALAKSHMI, M. ALEXE, A. PIGNOLET, N. D. ZAKHAROV, C. HARNAGEA, S. SENZ and D. HESSE, Appl. Phys. Lett. **74** (1999) 603.

[9] M. S. CHEN, T. B. WU, J. M. WU, Appl. Phys. Lett. **68** (1996) 1430.

[10] T. YU, Y. F. CHEN, Z. G. LIU, X. Y. CHEN, L. SUN, N. B. MING and L. J. SHI, Mater. Lett. **26** (1996) 73.

[11] G. D. HU, I. H. WILSON, J. B. XU, C. P. LI and S. P. WONG, Appl. Phys. Lett. **76** (2000) 1758.

[12] X. J. ZHANG, S. T. ZHANG, Y. F. CHEN, Z. G. LIU and N. B. MING, Microelectron. Eng. **66** (2003) 719

[13] K. M. SATYALAKSHMI, K. B. R. VARMA and M. S. HEGDE, J. Appl. Phys. **78** (1995) 1160.

[14] P. WANG, W. WU, G. HU, S. FAN, Y. DING and H. WU, Surf. Rev. Lett. **14** (2007) 123.

[15] G. D. HU, S. H. FAN and X. CHENG, J. Appl. Phys. **101** (2007) 054111.